\documentclass[runningheads]{llncs}
\usepackage[T1]{fontenc}
\usepackage{graphicx}
\usepackage{booktabs}
\usepackage{lscape}
\usepackage{multirow}
\usepackage{subcaption}
\usepackage{comment}

\usepackage{tikz}
\def\checkmark{\tikz\fill[scale=0.4](0,.35) -- (.25,0) -- (1,.7) -- (.25,.15) -- cycle;}

\begin{document}
\title{Towards the interoperability of low-code platforms}
%
%
\author{Iván Alfonso\inst{1}$^*$ \and
Aaron Conrardy\inst{1,2}$^*$ \and
Jordi Cabot\inst{1,2}}
\authorrunning{I. Alfonso et al.}

\def\thefootnote{*}\footnotetext{These authors contributed equally to this work}\def\thefootnote{\arabic{footnote}}
%
\institute{Luxembourg Institute of Science and Technology, Esch-sur-Alzette, Luxembourg \email{\{ivan.alfonso,aaron.conrardy,jordi.cabot\}@list.lu} \and
University of Luxembourg,  Esch-sur-Alzette, Luxembourg}
\maketitle              
\begin{abstract}
With the promise of accelerating software development, low-code platforms (LCPs) are becoming popular across various industries. Nevertheless, there are still barriers hindering their adoption. Among them, vendor lock-in is a major concern, especially considering the lack of interoperability between these platforms. Typically, after modeling an application in one LCP, migrating to another requires starting from scratch remodeling everything (the data model, the graphical user interface, workflows, etc.), in the new platform. 

To overcome this situation, this work proposes an approach to improve the interoperability of LCPs by (semi)automatically migrating models specified in one platform to another one. The concrete migration path depends on the capabilities of the source and target tools. We first analyze popular LCPs, characterize their import and export alternatives and define transformations between those data formats when available. This is then complemented with an LLM-based solution, where image recognition features of large language models are employed to migrate models based on a simple image export of the model at hand. The full pipelines are implemented on top of the BESSER modeling framework that acts as a pivot representation between the tools.


\end{abstract}

\section{Introduction}
\label{sec:intro}


Model-driven approaches such as model-driven engineering (MDE), model-driven architecture (MDA) or model-driven development (MDD) aim to streamline and optimize the delivery of applications from design through development, deployment, and operation by leveraging models at different levels of abstraction \cite{brambilla2017model}.

Building on these concepts, the term low-code has become popular in the software industry.  
Low-code can be seen as a pragmatic style of model-based approaches, where a fixed and reduced number of model types (enough to address the type of application being generated) are used together to model the system-to-be followed by a direct transformation of those models into the running system either via code-generation or model interpretation  \cite{cabot2024thelowcode}. More specifically, low-code platforms (LCPs) usually provide three types of (mostly graphical) modeling sublanguages: a data modeling language, a user interface (UI) language and a behavioural language linking the two, i.e. stating what actions (e.g. CRUD operations) on the data model should be triggered based on a UI event.

Analyses of the current market, such as Gartner's Magic Quadrant \cite{oleksandr2024magic}, assess and positions LCPs based on various criteria, such as their industry leadership, features, vision, etc. For example, LCPs like Mendix, OutSystems, and PowerApps are positioned as market leaders and are among the most widely used across different industries in areas such as process automation, financial services, public sector industries, and others. Unfortunately, all major players are commercial products with limited import and export capabilities creating a vendor lock-in effect, which represents a significant risk for users and companies \cite{yan2021impacts,bock2021low}.

In this paper, we study the interoperability between the emergent market of low-code tools as it can have a significant impact in the adoption and long-term success of this new model-based style of software development. Interoperability is defined as the ability of two or more software components to cooperate despite differences in language, interface, and execution platform \cite{wegner1996interoperability}. It is essential in scenarios that require collaborative work, cross-platform migration, forward and reverse engineering, system integration, and other similar tasks. 


We first analyze the current import/export options offered by major players in this market. Based on the identified limitations, we propose a model-based migration approach to enable interoperability among the leading LCPs. This model-based approach is enhanced with capabilities of visual large language models (LLMs) when needed.

The remainder of the paper is organized as follows:
In Section \ref{sec:motivation}, we perform the interoperability analysis of LCPs. Section \ref{sec:overview} offers an overview of our proposed solution. Section \ref{sec:export} details the export model options we propose while \ref{sec:import} details the process to import the exported models into a different tool. 
Section \ref{sec:tool} presents the tool support.
Section \ref{sec:roadmap} discusses some of the challenges we have identified. Finally, Section \ref{sec:related-work} reviews related work and Section \ref{sec:conclusion} concludes the paper.
\section{Evaluation of the interoperability among low-code tools}
\label{sec:motivation}

Vendor lock-in appears as the second most significant barrier in the adoption of low-code tools in a survey conducted by OutSystems \cite{outsystems2019}. The problem of vendor lock-in is also highlighted in other studies \cite{waqas2024using,sahay2020supporting}. Indeed, due to lack of interoperability among LCPs, when an application is developed on a specific LCP, migrating to another provider often requires redevelopment on the new platform \cite{sufi2023algorithms}.



Interoperability between two LCPs could be defined at the model or runtime level. The latter would imply that applications generated by one LCP could be executed on another LCP. The former implies that the model created in one LCP can be imported in another LCP to enable the generation of the application in that LCP without starting from scratch. We are interested in this second option as we do not consider the first one feasible given the different runtime libraries and dependencies employed by each LCP.

There is currently no study that focus on the import and export model capabilities of LCPs that would enable their interoperability. Therefore, we have evaluated ourselves the LCPs that are defined as Leaders or Challengers by Gartner's magic quadrant \cite{oleksandr2024magic}, as these are supposed to be the most popular and completes ones.

For each tool, we have evaluated the import and export capabilities for the three major types of models provided by each LCP. 


The collected data are summarized in Table \ref{tab:interoperability_evaluation}. Beyond marking which models can be imported (exported) we also indicate the format used to store the imported (exported) models. Obviously, we only consider formats that can be parsed to read the model. Proprietary or obfuscated file formats are not taken into account as they cannot be used in a migration pipeline.

Note that some tools require the installation of third-party solutions to export the models and/or offer limited import capabilities that derive partial schemas from existing data sources but cannot import explicit models. 

Particular emphasis was placed on analyzing the completeness of the data model during the export and import processes, given its central  role in a low-code process. For instance, some tools only offer limited support where only classes (but no associations) can be exported. 
This information is contained in the data model column of Table \ref{tab:interoperability_evaluation}, where exported/imported data models that lacked relationships were marked as "half". We would like to remark that more advanced data modeling constructs (like generalizations, association classes, enumerations,...) are completely ignored by the vast majority of the tools. 

\begin{table}[h]
    \centering
    \caption{Interoperability Evaluation of Low Code Platforms ("\protect\checkmark/2" symbolizes partial support, "\protect\checkmark$^*$" symbolizes the need for a 3rd-party application)}
    \label{tab:interoperability_evaluation}
    \begin{tabular}{|c||c|c|c|c|c|c|c|c|}
        \hline
        \multirow{2}{*}{\textbf{Platform}} & \multicolumn{4}{c|}{\textbf{Model Export}} & \multicolumn{4}{c|}{\textbf{Model Import}} \\  \cline{2-9}
        & \textbf{Data} & \textbf{GUI} & \textbf{Behav.} & \textbf{Format} &  \textbf{Data} & \textbf{GUI} & \textbf{Behav.}  & \textbf{Format} \\ 
       \hline
       \hline
        Mendix & \checkmark & \checkmark & \checkmark & JSON  & \checkmark/2 & & & XLSX \\
        \hline
         OutSystems & \checkmark$^{*}$ & & & XLSX  & \checkmark/2 & & & XLSX \\
         \hline
         PowerApps & \checkmark/2 & \checkmark & \checkmark & CSV + JSON  & \checkmark/2 & \checkmark & \checkmark & CSV + JSON \\
         \hline
         Appian & \checkmark & \checkmark & \checkmark & XML & \checkmark & \checkmark & \checkmark & XML \\
         \hline
         ServiceNow & \checkmark$^{*}$ & \checkmark$^{*}$ & \checkmark$^{*}$ &  XML  & \checkmark$^{*}$ & \checkmark$^{*}$ & \checkmark$^{*}$ &  XML\\
         \hline
         Salesforce & \checkmark$^{*}$ & & & XLSX   & \checkmark/2& &  &XLSX \\
         \hline
         Pegasystems & \checkmark &  & & XLSX  & \checkmark/2 & &  & XLSX \\
         \hline
         Zoho & \checkmark & \checkmark & \checkmark & DS &  \checkmark& \checkmark & \checkmark & XLSX + DS \\
         \hline
         ReTool & \checkmark/2 &\checkmark & \checkmark &  CSV + JSON  & \checkmark/2 &\checkmark&\checkmark  & CSV + JSON \\
         \hline
         Oracle Apex & \checkmark & \checkmark & \checkmark & SQL  & \checkmark & \checkmark & \checkmark & SQL \\
        \hline

    \end{tabular}
\end{table}

From this evaluation, we can draw several conclusions regarding the interoperability of LCPs, particularly in relation to vendor lock-in:

\begin{itemize}
    \item While some tools use the same format for storing their models (e.g. JSON) their internal schema is completely different therefore requiring a migration phase before they can be used in another tool
    \item Support for data models is (with limitations) common but there are many more limitations when it comes to the other types of models. In terms of options and in terms of diversity and quality of the import/export.
    \item Some tools do not offer a real model import. There is no option to pass them a model. But they offer the feature to initialize a project based on an existing data source ( a database, a CSV or Excel file,...). So a pseudo path to import a model in those tools is to create a, for instance, Excel file that somehow represents the model we want to import. Obviously, there is an information loss (e.g. ambiguity in the definition of the data types, absence of explicit relationships among classes, no validation rules...) as the model is not directly imported but has to be inferred from the CSV. For instance, Mendix, OutSystems, and PowerApps rely on this method. 
\end{itemize}

\section{Overview of our model-based migration approach}
\label{sec:overview}

\begin{figure}[t]
\centering
\includegraphics[width=1\textwidth]{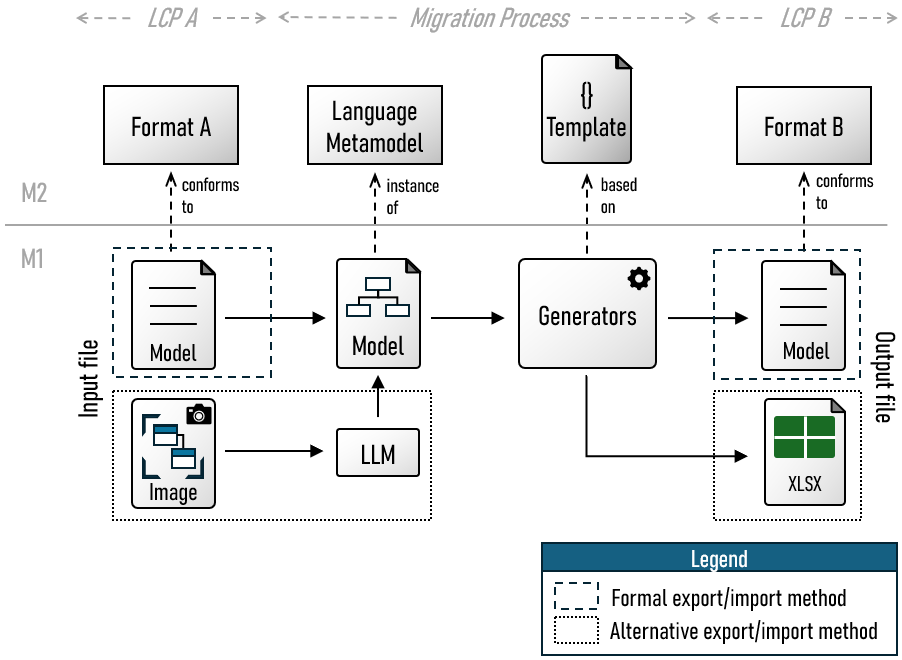}
\caption{Model-based migration solution for LCPs}
\label{fig:overview}
\end{figure}

Our model-driven solution for migrating models from an \textit{LCP A} to an \textit{LCP B} is summarized in Figure \ref{fig:overview}. 

The input file stores the application model/s. If the tool offers a formal model export feature, this file will contain the model expressed in a textual concrete syntax conforming to the tool language (e.g. a JSON file conforming to a JSON schema mapping the language concepts). Otherwise, we use as an alternative an image file capturing the graphical model (i.e. a screenshot of the model displayed in the tool). This dual input format accommodates LCPs that lack complete export support.

In both cases, the file is parsed and transformed into an intermediate model. In the first scenario (a formal textual description is available) a text-to-model transformation is used to create the model. This type of model transformation uses a grammar to parse the file and produces a model conforming to the language metamodel used as pivot representation between the two platforms. In the second scenario, an LLM is used to create the intermediate model directly from the provided image. 

These intermediate models conform to the B-UML metamodel, the modeling language we used as pivot language. B-UML, part of the  BESSER modeling framework \cite{alfonso2024building}, leverages and adapts well-known modeling standards, including UML for defining data models and IFML\footnote{\url{https://www.ifml.org/}} for UI models.


Once the \textit{Model} is created, it can be refined or completed, if needed. 

Finally, a template-based generator creates a textual representation of the model but now in terms of the concrete syntax expected by the target platform. When such platform does not offer a model import feature we rely on other available imports such as importing from Excel or CSV files which is common in most tools. Note that in this case we may lose some information as \textit{LCP B} is not directly importing the model but inferring it from the CSV structure. 


It is worth highlighting that, typically, model-driven interoperability approaches require using two metamodels to describe each tool’s internal schema together with a set of model-to-model transformations (M2M) to map the related concepts. Often, to avoid a combinatorial explosion problem when several tools need to be bridged, there is also an intermediate metamodel (B-UML) in our case. This approach is effective when there is a large semantic gap between the source and target tools and complex transformations need to be used to effectively transform the models to match the semantics of the target tool \cite{brambilla2017model}. While LCPs use different languages, they are close enough among them (i.e. most are variations of the core concepts you can find in standards like UML and IFML) that a more direct approach, like the one described in Figure \ref{fig:overview} offers a better trade-off. 

Note as well that being able to migrate models from \textit{LCP A} to \textit{LCP B} does not imply that the reverse is possible. For full interoperability we need to create both pipelines. And they may have different quality and/or require different combinations of the approaches described above depending on the features of each tool. 

Next sections cover these steps in more detail.
\section{Exporting the Model from LCP A}
\label{sec:export}

As discussed in Section \ref{sec:motivation}, not all LCPs support exporting a complete model or exporting models in a standardized textual format, and those that do typically use unique syntactic formats that differ across platforms. To address this challenge, we propose two methods for model migration (see Figure \ref{fig:overview}): (1) using the text file(s) generated by the official export feature of the LCP, or (2) an alternative solution providing a picture of the model to import, possibly combined with textual files containing an incomplete model.

\subsection{Formal Export Method: Text to model}\label{sec:export-formal}
\label{sec:export-t2m}
When the LCP provides functionality to export the model in a readable text-based format, the model's information becomes accessible and can be extracted to prepare it for migration to the second LCP. This initial step in the migration pipeline is carried out through a Text-to-Model (T2M) transformation. The T2M transformation, guided by specific transformation rules, extracts information about the model elements such as classes, attributes, and relationships for data models; buttons, layouts, and styles for GUI models; and workflows and events for behavioral models. It then constructs a new LCP-independent model, conforming to the B-UML modeling language.

The exact set of transformation rules will vary depending on each LCP and the syntax it uses, though they remain all similar. For instance, tools such as Mendix, Appian, or Apex enable the generation of text files containing data model information. For every tool we will have rules to identify in the input file the classes, attributes, associations of the model and create the corresponding elements in B-UML. 

\begin{figure}[t]
\centering
\includegraphics[width=1\textwidth]{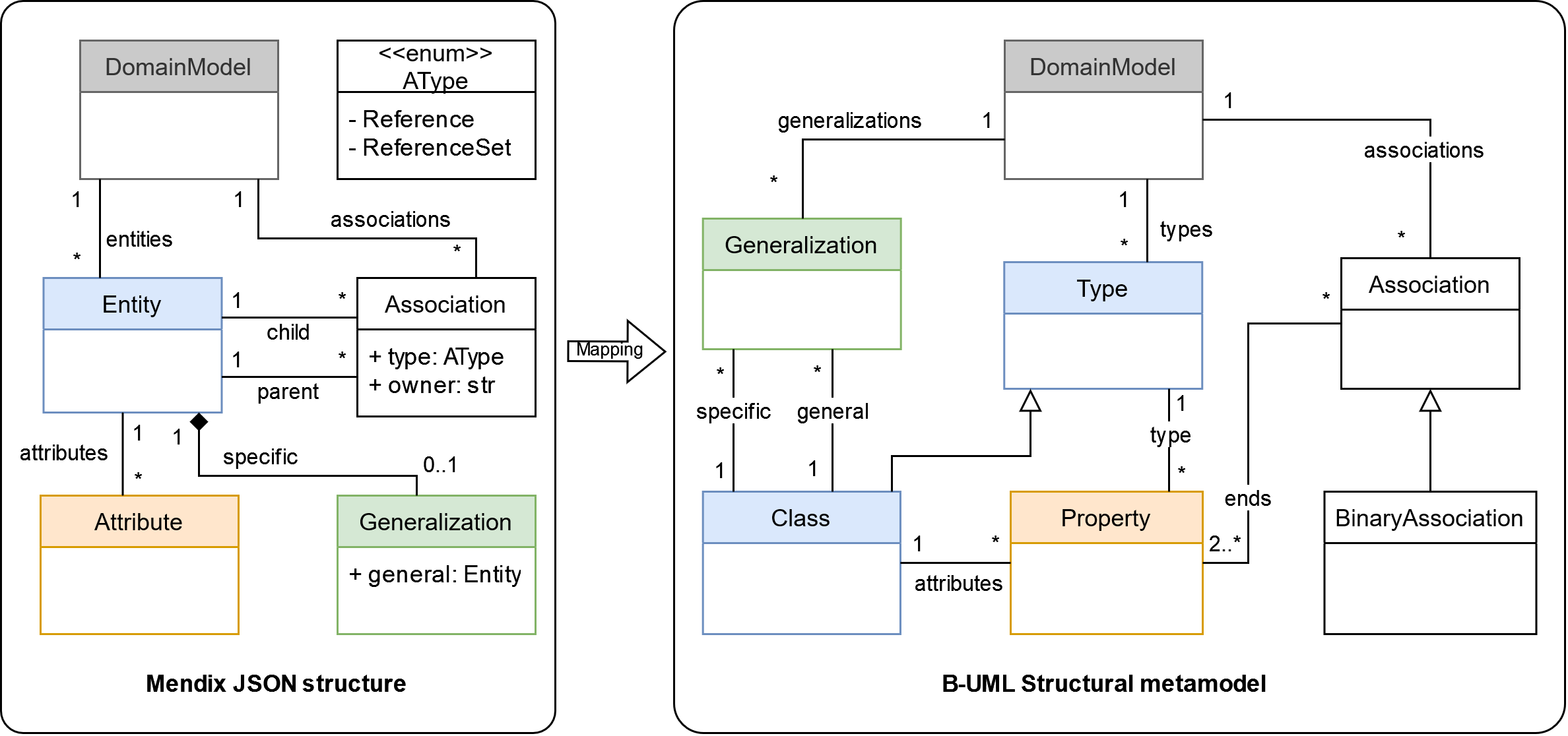}
\caption{Mapping of main concepts between Mendix and B-UML. Corresponding concepts are indicated by matching colors; for example, an \textit{Attribute} in Mendix is mapped to a \textit{Property} in B-UML}
\label{fig:mapping}
\end{figure}

As an example, we have completely implemented the transformation rules for Mendix data models (see Section \ref{sec:tool}), although, as we just said, the export process for other tools would be similar. Mendix allows exporting the full project as a JSON file using its command line tool. Given this JSON, our text-to-model transformation, implemented in Python, reads the JSON object and constructs a data model instance of the B-UML metamodel following the mapping rules between Mendix and B-UML depicted in Figure \ref{fig:mapping}.
Note how several of these concepts are directly transformed without significant modifications, such as \textit{DomainModel}, \textit{Entity}, \textit{Attribute}, and \textit{Generalization}. For \textit{Association}s, the ends are determined by the \textit{child} and \textit{parent} relationships, while the cardinality is defined by the \textit{type} and \textit{owner} attributes. Other concepts, such as \textit{PrimitiveDataType}s and \textit{Enumeration}s, are also transformed in a straightforward manner but are omitted from Figure \ref{fig:mapping} for simplicity.

\subsection{Alternative Export Method: Image to model}\label{sec:export-informal}
In the case that an LCP does not provide an export functions, we leverage the power of visual LLMs to parse the model depicted in an image taken from the source LCP. 

Indeed, previous efforts have already highlighted the capabilities of visual LLMs to transform images of data models to actual models \cite{img2uml}, especially when combining OpenAI's GPT as LLM, PlantUML\footnote{PlantUML is a popular textual syntax for UML models} as the target metamodel and the prompt "Can you turn this hand-drawn UML class diagram into the corresponding class diagram in PlantUML notation?".

Yet, for our purposes, a limitation of that previous study is the focus on data models defined as UML class diagrams, while each LCP provide its own data modeling language. Nevertheless, given that all those data modeling languages are heavily inspired by the typical data modeling primitives available in the UML (or ER) language, we have adapted the method to deal with LCP data modeling languages. 

The adaptation consisted in extending the previously mentioned prompt with additional context specific to each LCP. This context includes a description of the syntax of the graphical model of that particular LCP and other relevant information about the LCP language. When this alternative is used to complement a partial model export using the technique in Section \ref{sec:export-t2m}, we provide as well the partial model as additional resource and ask the LLM to respect the elements in that file and complement them with the missing ones (typically, associations that can be seen in the image but are missing in the export). 

\begin{figure}[t]
    \centering
    \includegraphics[width=\linewidth]{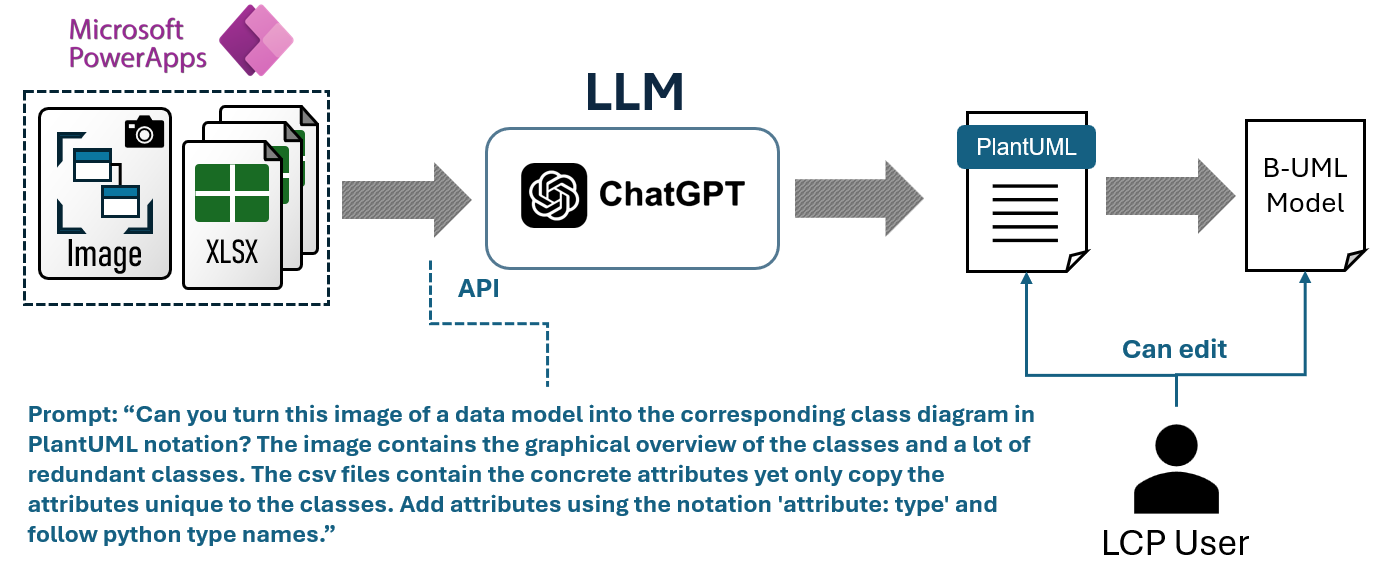}
    \caption{Overview of image to model pipeline with PowerApps data model}
    \label{fig:llm}
\end{figure}

An example image, textual file and prompt combination is given in Figure \ref{fig:llm}, where we used a data model created in PowerApps combined with the CSV files it provided as partial export (including the structure of classes but not the relationships among them). Note that since the LLM generates PlantUML models, we additionally need to transform PlantUML models to B-UML ones. This is a feature already available in B-UML. We could ask the LLM to directly generate B-UML models but as PlantUML is a format that the LLM has seen much more often in the training data, we get better results going through PlantUML

As the usage of an LLM does not guarantee the completeness and correctness of the transformation due to its nondeterministic nature, LCP users have the possibility to edit either the PlantUML or B-UML model to improve the results if necessary.

\section{Generating and Importing the Model to LCP B}
\label{sec:import}

In MDE, code generation refers to the process of transforming models into source code (M2T) \cite{brambilla2017model}. Our approach proposes generators to transform a B-UML model into the file(s) that represent the model in a format readable and importable by the target low-code platform (\textit{LCP B} in Figure \ref{fig:overview}).


We address generators for two scenarios: (1) generating the model for import into an LCP with a direct model importing functionality, and (2) producing a spreadsheet (XLSX file) as an alternative solution for LCPs not offering a model import feature but able to infer models from structured data sources such as XLSX files. 

\subsection{Formal Import Method}
\label{subsec:formal-import}

When an LCP provides a formal importer for models, the corresponding model file can be generated according to the LCP experted format via a M2T (Model-to-Text) transformation. This M2T transformation must map the B-UML concepts to the concepts of the LCP language and write them using the right concrete syntax. Basically, it is the reverse process to the one we saw in Section \ref{sec:export-formal}. For example, Appian and Oracle Apex allow the import of the entire project, including the GUI, behavior, and data models definition from a set of XML and SQL files. 


As an example of an M2T transformation, we have developed a generator for Oracle Apex. This generator produces specific statements to define the data model in Oracle Apex, which stores data models as relational databases and UI and behvaioral models as rows in a special set of predefined tables. The transformation rules for the data part of the M2T process were derived from \cite{daniel2019umlto}, which defines the mapping of UML class diagrams to relational databases. An example of this generator is presented in Section \ref{sec:tool}.

\subsection{Alternative Import Method}
\label{subsec:informal-import}

Most low-code platforms (LCPs) provide methods to import existing data in various formats, such as XLSX, JSON, and CSV as way to initialize a new modeling project. 
As part of the initialization process, when the data is ingested from these external files, the platform infers a data model to be used to manage the imported data.
For example, when importing data with a XLSX file, an LCP generates a data model where for each sheet the LCP creates a new class, for each column an attribute of that class, and the cells' format of the column is used as a data type. Some LCPs can recognize and generate also relationships between classes when there is a linked field in the sheets, but this is not common.

While this approach is not a formal or standardized method for importing data models (as it is primarily designed for data import), we propose using it as an alternative for platforms that lack a dedicated mechanism for data model import. Therefore, we have developed a generator that produces a structured spreadsheet file representing the input model’s structure. The generator adheres to the following rules:

\begin{itemize}
    \item A \textit{DomainModel} generates a single spreadsheet.
    \item Each \textit{Class} generates a separate sheet named after the class.
    \item Each attribute (\textit{Property}) of a class generates a new column in the respective sheet. The column is formatted according to the data type of the Property. For instance, if the data type is "date," the column is formatted with the style "DD/MM/YYYY".
    \item Associations with cardinality many-to-one generate a new column in one sheet, where the column cells are configured to display a drop-down list referencing the associated sheet (the other Class in the association).
    Many-to-many associations generate an additional sheet (bridge sheet) containing two columns that link the sheets (Classes) involved in the association.
    \item Finally, an example row of data is generated with default values that respect each cell data format. As mentioned, importing a spreadsheet is typically a method for ingesting data rather than a data model. Therefore, the LCP requires at least one data entry to recognize data types and relationships between sheets.
\end{itemize}

Although this is an alternative for importing the data model, it is not a valid alternative for UI and behavioral models. Additionally, since a spreadsheet with data is not truly as semantically rich as models themselves going through a  spreadsheet representation may result in a model of lower quality that may need to be refined a posteriori.


\section{Tool support}
\label{sec:tool}

As proof of concept, we have implemented some migration paths on top of the BESSER platform. Our prototype is open-source and available in the project repository\footnote{https://github.com/BESSER-PEARL/BESSER-Migration-Hub.git}. The implemented paths exercise the different import/export combinations described in the previous sections. Our current tool support is restricted to data models and support for the interoperability of UI and behavioural models would follow the directives explained so far but it's left as future work.


BESSER is an open-source low-code platform implemented in Python, featuring a modeling framework and a language called B-UML. This language supports the specification of several model types, including structural (or data), object, GUI, deployment, and state machines, adhering to recognized standards such as UML and IFML. In our implementation, models intended for migration are transformed into B-UML models that can still be edited to complete the imported model before regenerating it according to the target LCP.



\begin{figure}
    \centering
    \includegraphics[width=0.8\linewidth]{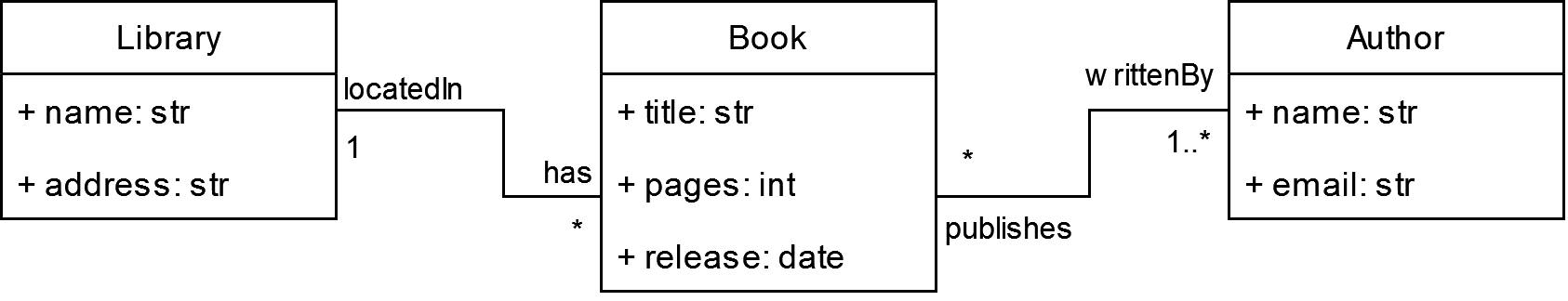}
    \caption{Library UML class diagram}
    \label{fig:libraryclassdiagram}
\end{figure}

As an example, we show the transformation of a simple library data model (see the UML class diagram for this model in Figure \ref{fig:libraryclassdiagram}). This model is originally defined in Mendix. Let's assume we want to first export it from Mendix and import it into PowerApps. 

\begin{figure}[ht]
    \centering
    \begin{subfigure}[b]{0.50\textwidth}
        \centering
        \includegraphics[width=\textwidth]{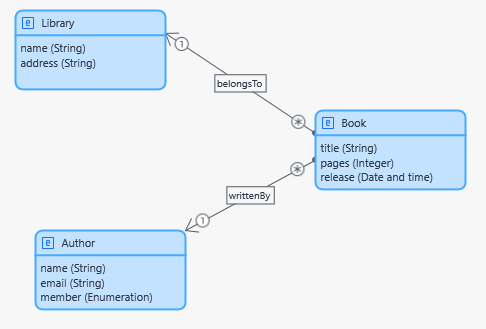}
        \caption{Initial data model in Mendix}
    \end{subfigure}
    \hfill
    \begin{subfigure}[b]{0.49\textwidth}
        \centering
        \includegraphics[width=\textwidth]{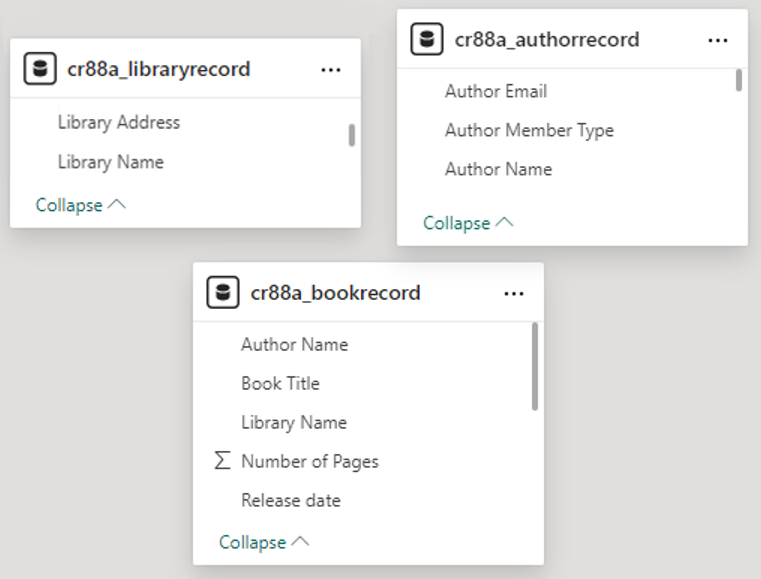}
        \caption{Resulting data model in PowerApps}
    \end{subfigure}

    \caption{From formal export to alternative import method}
    \label{fig:examples1}
\end{figure}

First, the Mendix data model is exported as a JSON file using the formal method provided by Mendix. Then, a T2M transformation for Mendix takes the Mendix model and transforms it into a B-UML compliant version. 
To be imported into PowerApps, the B-UML model is given as input to the generator for Excel files that will produce a spreadsheet that PowerApps will be able to import and use to recreate the model. 
We see the result in Figure \ref{fig:examples1}.
Although the generated Excel file includes associations through columns linking different sheets, due to limitations on PowerApps' side, the model is not complete as it is missing its relationships so this part needs to be manually completed. 

\begin{figure}[ht]
    \centering
    \begin{subfigure}[b]{0.60\textwidth}
        \centering
        \includegraphics[width=\textwidth]{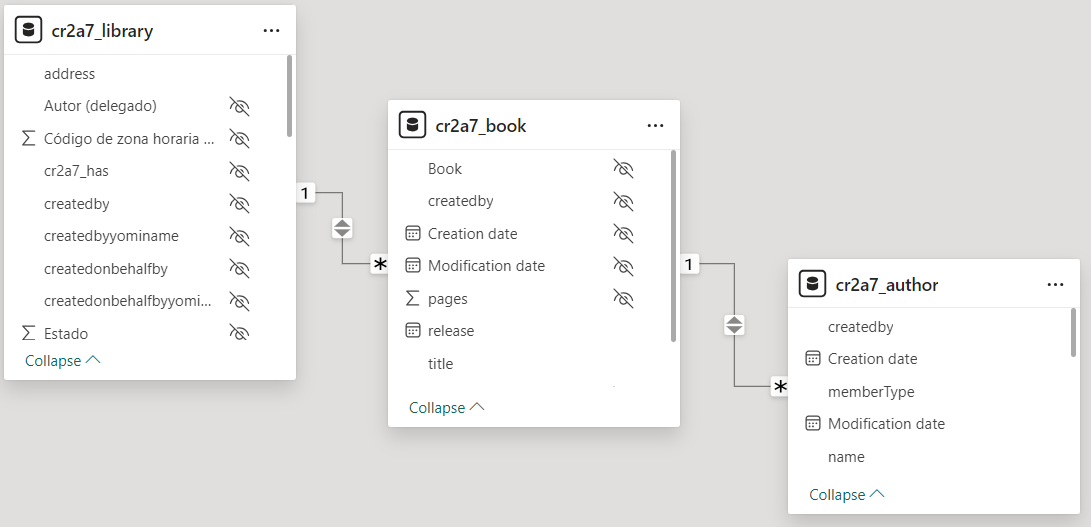}
        \caption{Initial model in PowerApps}
    \end{subfigure}
    \hfill
    \begin{subfigure}[b]{0.30\textwidth}
        \centering
        \includegraphics[width=\textwidth]{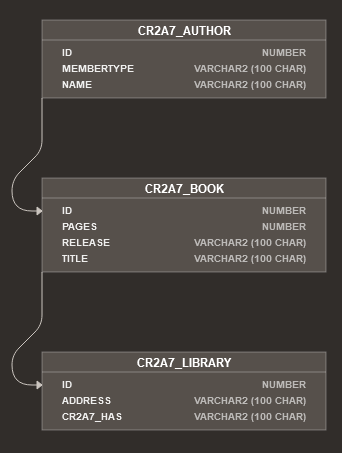}
        \caption{Resulting data model in Oracle Apex}
    \end{subfigure}

    \caption{From alternative export to formal import method}
    \label{fig:examples2}
\end{figure}

As a second example, assume that now we want to move from PowerApps to Oracle Apex. In this case, we use a screenshot of the PowerApps data model and the CSV files that PowerApps is able to export representing the classes (but not the associations) and we fed both to the LLM combined with a PowerApps specific prompt as described in \ref{sec:export-informal}. As before, this process produces a B-UML model.  This model is then used to generate a semantically equivalent Oracle SQL database structure (using the generator presented in Section \ref{subsec:formal-import}), which is then read by Oracle Apex. We see the result in Figure \ref{fig:examples2}.

In both cases, some elements are missing or wrong as a result of the exporting/importing limitations of the demonstrated LCPs. 
Regardless, the data model has been successfully transferred and requires little completion by the user.
Depending on the LCPs involved in the process, the completion steps can either take place at the B-UML level, in the LCP itself or not be required at all.

\section{Discussion}
\label{sec:roadmap}


This section discusses some findings and reflections on the interoperability challenge.

\subsubsection{Lack of support for import/export models in LCPs}\label{subsec:export_import}
We found support for import and export of models rather limited. As seen in section \ref{sec:motivation}, some tools offer very limited options, often relying on restricted CSV files for data models only. And even for those options, in some tools, finding the way to do it was "hidden" in the docs. Clearly, facilitating users to move out to a different LCP is not a priority and, instead, vendor lock-in is a reality. 

Additionally, there seems to be a clear lack of interest in advancing model migration capabilities, likely because most LCPs are commercial platforms with little incentive to enable the export of their projects to other platforms. In some cases, even finding clear documentation for exporting models was not trivial.



\subsubsection{Lack of standard interchange model}

The lack of a standard interchange model for LCPs is not helping either as each LCP follows their own format using arbitrary file types and internal schemas, increasing interoperability issues. This had been also a problem for UML tools until the UML diagram interchange specification\footnote{\url{https://www.omg.org/spec/UMLDI/1.0/About-UMLDI}} was proposed to allow for a seamless interchange between tools supporting the development of UML models.
A similar endeavor would be required for LCP models, enabling a seamless migration of applications.

Yet, compared to the UML DI specification, we believe that more effort would be required. 
Firstly, while LCP language are similar, there are still various differences in the supported modeling features. 
Secondly, the addition of GUI models in LCPs drastically increases the difficulty of developing such a standard interchange model.
We experienced the former while using the B-UML language as a candidate standard model to unify the LCP's data models.
Indeed, the inconsistencies among available modeling features makes the creation of a common language difficult.
Nevertheless, the creation of a standard interchange model for LCPs might lead to vendors addressing the inconsistencies. If they at least could agree on a standard for data models (the more similar among them) that would be already helpful. Ideally, the interchange format should also include details to replicate the model layout.

\subsubsection{Automatic model recovery techniques}
Limitations in the import/export features may result in incomplete models. While users can manually complete the models, it would be useful to propose model recovery techniques able to autocomplete the partial models. This autocompletion could be done with AI techniques or simpler heuristic-based rules \cite{cabot2024lowmodeling}. This autocompletion could go beyond single models and even aim to build basic UI and behavioral models from the data model so that users do not need to start from scratch with those models when they cannot be imported. For instance, a CRUD-based IFML-like model can be automatically derived from a data model \cite{Rodriguez-Echeverria18}. 


\subsubsection{Incremental migration}
Beyond importing entire models, LCPs should be able to recognize existing components and only append new elements to existing models or update existing ones. Moreover, this update should respect changes done by the user on those elements instead of overwriting (similar to the concept of protected areas in code-generation approaches).  

\subsubsection{More advanced, AI-based, alternative import and export methods}
Additional alternative export or import methods leveraging AI could be explored.
For the former, multi-image or video LLM-based processing capabilities might be necessary to adequately capture and understand large models created in a LCP without a proper exporting feature. 
 For the import, a potential solution would be to train an AI system how to replicate an input model using the target LCP interface for LCPs without proper importing capabilities. LLMs such as Claude\footnote{https://docs.anthropic.com/en/docs/build-with-claude/computer-use} recently proposed a \textit{computer use} feature that can be used to manipulate a computer desktop environment. This opens the door to teach such systems how a human would replicate in a target platform the model that is displayed in the source one. The idea would be that after replicating a small number of elements, the LLM would be able to understand the process and replicate it for the rest of the model.

\section{Related work}
\label{sec:related-work}

Several literature review studies \cite{bucaioni2022modelling,pinho2023usability,alsaadi2021factors} have focused on analyzing the low-code development ecosystem. These studies highlight the concern of vendor lock-in in low-code platforms (LCPs) as a significant dependency risk for customers. However, while these studies identify the vendor lock-in problem, they neither analyze the interoperability aspects of the platforms nor propose solutions to address this issue.

A more detailed comparative analysis of several LCPs is presented in \cite{sahay2020supporting}. This study evaluates various features, including security, scalability, deployment support, and interoperability. Its interoperability analysis focuses on determining whether the LCP supports interaction with external services (such as Dropbox or SharePoint) and connectivity to external data sources for building forms. However, the authors do not explore specific limitations regarding application component migration, such as the feasibility of migrating models between LCPs. Additionally, proposing or designing a solution to address this problem is beyond the scope.

While these studies primarily focus on evaluating interoperability at a high level, other research has explored technical solutions to ensure interoperability between modeling platforms using MDE, often referred to as bridges. The most studied cases in the literature propose bridges between the Eclipse Modeling Framework (EMF) and other technologies. For example, solutions have been developed for EMF and Visio \cite{kern2009integration}, EMF and ADOxx \cite{bork2022towards}, EMF and MSDKVS \cite{cesal2024establishing}, as well as EMF and Microsoft DSL Tools \cite{bruneliere2010towards}. However, these solutions assume that the models involved are accessible and serializable (e.g., EMF provides models in XMI format), an assumption that does not apply to all LCPs. In contrast, some LCPs analyzed in our study lack mechanisms to export or import models in a standard format. Furthermore, since LCPs are closed-source, it becomes challenging to develop extensions to support such functionality. Our solution is inspired by these previous ones but adapted (and extended, e.g. via LLMs) to deal with the specificities of the LCP ecosystem. 

The idea of using a language as pivot representation has also been explored in the past. For example, the unified metamodel KF proposed in \cite{fahad2008er2owl} integrates the concepts of ER, ORM2, and UML, along with a set of transformation rules to map these models to KF. Additionally, other approaches focus on transforming UML class diagrams into OWL ontologies \cite{vo2020transformation} or converting OWL to ER models \cite{fahad2008er2owl}. However, these solutions rely on source models that are built using standardized languages such as UML. In the case of LCPs, data models often do not adhere to any standard modeling language. In some cases, it is necessary to resort to more generic and flexible solutions, including as mentioned before, the need to deal with partial models. 

To the best of our knowledge, ours is the first study to propose a solution to address the vendor lock-in problem of LCPs, which are becoming increasingly adopted. Our solution not only proposes bridge transformations to enable model export and import between LCPs but also introduces alternative methods using an LLM to migrate data models from a screenshot of their graphical representation.

\section{Conclusion}
\label{sec:conclusion}

In this paper, we propose an approach to improve the interoperability of LCPs by (semi-)automatically migrating models from one platform to another. Depending on the capabilities of the LCP to import and export models, we propose different interoperability pipelines. 

As further work, we want to tackle the challenges outlined in Section \ref{sec:roadmap}. Additionally, we will extend our tool support, especially regarding the use of LLMs to export UI and behavioural models, to enable more interoperability bridges among the tools that cover not only the data model but also this behavioural and UI models.  Moreover, also as part of the tool support, we plan to develop a graphical user interface to facilitate the use of our tool. LCP users will be able to indicate the source and target LCPs together with the input files (including screenshots if needed) and automatically obtain the ready-to-import files. Depending on the migration path, they may be prompted to answer a few questions to optimize the quality of the result (e.g. to validate the model inferred by a visual LLM).

\bibliographystyle{splncs04}
\bibliography{biblio}

\begin{thebibliography}{10}
\providecommand{\url}[1]{\texttt{#1}}
\providecommand{\urlprefix}{URL }
\providecommand{\doi}[1]{https://doi.org/#1}

\bibitem{alfonso2024building}
Alfonso, I., Conrardy, A., Sulejmani, A., Nirumand, A., Ul~Haq, F., Gomez-Vazquez, M., Sottet, J.S., Cabot, J.: Building besser: an open-source low-code platform. In: International Conference on Business Process Modeling, Development and Support. pp. 203--212. Springer (2024)

\bibitem{alsaadi2021factors}
Alsaadi, H.A., Radain, D.T., Alzahrani, M.M., Alshammari, W.F., Alahmadi, D., Fakieh, B.: Factors that affect the utilization of low-code development platforms: survey study. Romanian Journal of Information Technology \& Automatic Control/Revista Rom{\^a}n{\u{a}} de Informatic{\u{a}} și Automatic{\u{a}}  \textbf{31}(3) (2021)

\bibitem{bock2021low}
Bock, A.C., Frank, U.: Low-code platform. Business \& Information Systems Engineering  \textbf{63},  733--740 (2021)

\bibitem{bork2022towards}
Bork, D., Anagnostou, K., Wimmer, M.: Towards interoperable metamodeling platforms: the case of bridging adoxx and emf. In: International Conference on Advanced Information Systems Engineering. pp. 479--497. Springer (2022)

\bibitem{brambilla2017model}
Brambilla, M., Cabot, J., Wimmer, M.: Model-Driven Software Engineering in Practice, 2nd Edition. Synthesis Lectures on Software Engineering, Morgan {\&} Claypool Publishers (2017)

\bibitem{bruneliere2010towards}
Bruneliere, H., Cabot, J., Clasen, C., Jouault, F., B{\'e}zivin, J.: Towards model driven tool interoperability: Bridging eclipse and microsoft modeling tools. In: European Conference on Modelling Foundations and Applications. pp. 32--47. Springer (2010)

\bibitem{bucaioni2022modelling}
Bucaioni, A., Cicchetti, A., Ciccozzi, F.: Modelling in low-code development: a multi-vocal systematic review. Software and Systems Modeling  \textbf{21}(5),  1959--1981 (2022)

\bibitem{cabot2024thelowcode}
Cabot, J.: The low-code handbook: Learn how to unlock faster and better software development with low-code solutions (10 2024)

\bibitem{cabot2024lowmodeling}
Cabot, J.: Low-modeling of software systems. In: Fill, H.G., Dom{\'i}nguez~Mayo, F.J., van Sinderen, M., Maciaszek, L.A. (eds.) Software Technologies. pp. 19--28. Springer Nature Switzerland, Cham (2024)

\bibitem{cesal2024establishing}
Cesal, F., Bork, D.: Establishing interoperability between emf and msdkvs: an m3-level-bridge to transform metamodels and models. Software and Systems Modeling pp. 1--30 (2024)

\bibitem{img2uml}
Conrardy, A.D., Cabot, J.: From image to {UML:} first results of image-based {UML} diagram generation using llms. In: Proceedings of the {STAF} 2024 Workshops: AgileMDE 2024, {LLM4MDE} 2024, and MeSS 2024 co-located with the International Conference on Software Technologies: Applications and Foundations {(STAF} 2024) Enschede, The Netherlands, July 8-11, 2024. {CEUR} Workshop Proceedings, vol.~3727, pp. 55--65. CEUR-WS.org (2024)

\bibitem{daniel2019umlto}
Daniel, G., G{\'o}mez, A., Cabot, J.: Umlto [no] sql: mapping conceptual schemas to heterogeneous datastores. In: 2019 13th International Conference on Research Challenges in Information Science (RCIS). pp. 1--13. IEEE (2019)

\bibitem{fahad2008er2owl}
Fahad, M.: Er2owl: Generating owl ontology from er diagram. In: International Conference on Intelligent Information Processing. pp. 28--37. Springer (2008)

\bibitem{kern2009integration}
Kern, H., K{\"u}hne, S.: Integration of microsoft visio and eclipse modeling framework using m3-level-based bridges. In: Proceedings of Second Workshop on Model-Driven Tool and Process Integration (MDTPI) at ECMFA, CTIT Workshop Proceedings. pp. 13--24. Citeseer (2009)

\bibitem{oleksandr2024magic}
Oleksandr, M., Kyle, D., Akash, J.: Magic quadrant for enterprise low-code application platforms. Gartner report  (2024)

\bibitem{outsystems2019}
OutSystems: The state of application development: Is it ready for disruption? (2019)

\bibitem{pinho2023usability}
Pinho, D., Aguiar, A., Amaral, V.: What about the usability in low-code platforms? a systematic literature review. Journal of Computer Languages  \textbf{74},  101185 (2023)

\bibitem{Rodriguez-Echeverria18}
Rodr{\'{\i}}guez{-}Echeverr{\'{\i}}a, R., Preciado, J.C., Sierra, J., Conejero, J.M., S{\'{a}}nchez{-}Figueroa, F.: Autocrud: Automatic generation of {CRUD} specifications in interaction flow modelling language. Sci. Comput. Program.  \textbf{168},  165--168 (2018)

\bibitem{sahay2020supporting}
Sahay, A., Indamutsa, A., Di~Ruscio, D., Pierantonio, A.: Supporting the understanding and comparison of low-code development platforms. In: 2020 46th Euromicro Conference on Software Engineering and Advanced Applications (SEAA). pp. 171--178. IEEE (2020)

\bibitem{sufi2023algorithms}
Sufi, F.: Algorithms in low-code-no-code for research applications: a practical review. Algorithms  \textbf{16}(2), ~108 (2023)

\bibitem{vo2020transformation}
Vo, M.H.L., Hoang, Q.: Transformation of uml class diagram into owl ontology. Journal of Information and Telecommunication  \textbf{4}(1),  1--16 (2020)

\bibitem{waqas2024using}
Waqas, M., Ali, Z., S{\'a}nchez-Gord{\'o}n, M., Kristiansen, M.: Using lowcode and nocode tools in devops: A multivocal literature review. New Perspectives in Software Engineering pp. 71--87 (2024)

\bibitem{wegner1996interoperability}
Wegner, P.: Interoperability. ACM Computing Surveys (CSUR)  \textbf{28}(1),  285--287 (1996)

\bibitem{yan2021impacts}
Yan, Z.: The impacts of low/no-code development on digital transformation and software development. arXiv preprint arXiv:2112.14073  (2021)

\end{thebibliography}

\end{document}